\documentclass[aps,pre,preprint,superscriptaddress,amsmath,amssymb,nofootinbib]{revtex4}

\usepackage{amsmath}
\usepackage{graphicx}
\usepackage{amssymb}
\usepackage{dcolumn}
\usepackage{bm}
\usepackage{color}
\usepackage{float}
\usepackage{upgreek}
\usepackage{gensymb}
\usepackage[final]{hyperref} 
\hypersetup{
colorlinks=true,       
linkcolor=blue,        
citecolor=blue,        
filecolor=magenta,     
urlcolor=blue         
}

\usepackage{upgreek}
\usepackage[normalem]{ulem}

\makeatletter

\begin{document}
\title{Diffusion of chiral active particles in a Poiseuille flow}
	
\author{Narender Khatri}\thanks{narenderkhatri8@iitkgp.ac.in}
\affiliation{Department of Physics, Indian Institute of Technology Kharagpur, Kharagpur - 721302, India}

\author{P. S. Burada}\thanks{Corresponding author: psburada@phy.iitkgp.ac.in}
\affiliation{Department of Physics, Indian Institute of Technology Kharagpur, Kharagpur - 721302, India}

\date{\today}
	
\begin{abstract} 
 We study the diffusive behavior of chiral active (self-propelled) Brownian particles in a two-dimensional microchannel with a Poiseuille flow. 
Using numerical simulations, we show that the behavior of the transport coefficients of particles, for example, the average velocity $v$ and the effective diffusion coefficient $D_{eff}$, strongly depends on flow strength $u_0$, translational diffusion constant $D_0$, rotational diffusion rate $D_\theta$, and chirality of the active particles $\Omega$.
It is demonstrated that the particles can exhibit upstream drift, resulting in a negative $v$, for the optimal parameter values of $u_0$, $D_\theta$, and $\Omega$. 
Interestingly, the direction of $v$ can be controlled by tuning these parameters. 
We observe that for some optimal values of $u_0$ and $\Omega$, the chiral particles aggregate near a channel wall, and the corresponding $D_{eff}$ is enhanced. 
However, for the nonchiral particles ($\Omega = 0$), the $D_{eff}$ is suppressed by the presence of Poiseuille flow.
 It is expected that these findings have a great potential for developing microfluidic and lab-on-a-chip devices for separating the active particles. 
\end{abstract}
	
\maketitle

\section{Introduction}

During the past decades, the dynamics of self-propelled particles in the low Reynolds number regime have attracted a lot of interest from the biology and physics communities \cite{Ramaswamy_ARCMP, Cates_RPP, Ebeling_EPJST, Lowen_RMP}.
Biological organisms, e.g., bacteria, sperm cells, ciliated microorganisms, etc., can swim independently without the need for external forces \cite{Brown_Nature, Berg_Nature, Goldstein_PRL,Howard_Sc, Woolley_Reproduction, friedrich}. 
The dynamics of these organisms occur at the low Reynolds number regime, i.e., the viscous force due to the surrounding fluid medium is dominant than the inertia that arises due to their mass \cite{purcell}. 
These self-propelled particles employ different propulsion mechanisms to move in a fluid, e.g., ciliated microorganisms swim with the help of the metachronal waves generated by the synchronous beating of cilia \cite{lighthill, blake}, sperm cells swim with the help of flagella \cite{friedrich}, and bacteria propel using a run and tumble mechanism \cite{patteson}. 
Motivated by the motion of biological organisms, researchers developed artificial self-propelled particles, micro- and nanoparticles, to understand the collective behavior of the living microorganisms as well as to use them for technological applications 
\cite{Jiang_Book, Muller_CR}. 
However, the swimming mechanism of artificial self-propelled particles is different from that of real microorganisms.
For example, the artificial particles consume energy from the external  energy source and transform it under nonequilibrium conditions into directed motion \cite{Ramaswamy_ARCMP}.
Typically, these artificial self-propelled particles, e.g., Janus particles, consist of two distinct faces, out of which only one is chemically active \cite{Howse_PRL}.
Remarkably, because of their functional asymmetry, these particles can induce either concentration gradients (self-diffusiophoresis) by catalyzing a chemical reaction on their active surface \cite{Howse_PRL, Paxton, Gibbs_APL, Volpe_SM} or thermal gradients (self-thermophoresis) by inhomogeneous light absorption \cite{Sano_PRL}. 
Note that the understanding of these artificial self-propelled particles can provide numerous promising applications in several fields, e.g., drug delivery through tissues, localize pollutants in soils, nonequilibrium self-assembly, etc. \cite{Weibel_PNAS, Chin_Lab, Yang_SM}.  

The diffusion of self-propelled particles in confined structures is of great interest \cite{Ghosh_PRL, Ao_EPL, Ai_PRE}. 
When compared to passive Brownian particles, active particles diffusing in confined structures exhibit peculiar features, resulting  for example, in spontaneous rectification \cite{Ghosh_PRL, Ai_PRE, Angelani_PRL}, phase separation of particles \cite{Speck_PRL, Fily_PRL}, collective motion in complex systems \cite{Vicsek_PR, Vicsek_Nature}, spiral vortex formation in circular confinement \cite{Dunkel_PRL}, depletion of elongated particles from low-shear regions \cite{Stocker_Nature}, trapping of particles in microwedge \cite{Kaiser_PRL}, and many more.
Note that the behavior of active particles in microchannels with a Poiseuille flow has been investigated both experimentally \cite{Hill_et_al@PRL:2007, Kaya_Koser, Kantsler_et_al@Elife:2014, Jing_et_al@SA:2020} and theoretically \cite{Jing_et_al@SA:2020, Zottl_Stark@PRL:2012, Ezhilan_Saintillan@JFM:2015, Vennamneni_et_al@JFM:2020, Costanzo_JPCM, Nash_PRL, Stark_EPJE, Stark_EPJSP, Shendruk_PRL, Peng_PRF, Chilukuri_PF}, reporting the upstream flow of particles.
Thus, the upstream flow is ubiquitous to active particles in a Poiseuille flow.
Ezhilan and Saintillan \cite{ Ezhilan_Saintillan@JFM:2015} studied the effect of Poiseuille flow on the transport of nonchiral slender active Brownian particles. 
Using numerical and asymptotic solutions to the Smoluchowski equation, they systematically investigated the distribution of particles in the channel and predicted net upstream flow. 
On the other hand, Peng and Brady \cite{Peng_PRF} formulated the transport of nonchiral active Brownian particles in a Poiseuille flow from a continuum perspective using the Smoluchowski equation.
Using different numerical methods, they calculated the transport properties, i.e., the average velocity and effective diffusion coefficient, and particles distribution in the channel from an effective advection-diffusion equation.
Recently, Jing et al. \cite{Jing_et_al@SA:2020} studied the chirality-induced bacterial rheotaxis in a Poiseuille flow using a combined experimental, numerical, and theoretical analysis.
They systematically investigated the average rheotactic velocity, velocity distribution, and orientation distribution of \textit{E. coli} bacteria in a channel flow.
However, the transport properties and distribution of chiral active Brownian particles in a channel with the Poiseuille flow are not fully understood and quantitatively scrutinized, which have a great potential in  
designing microfluidic and lab-on-a-chip devices to control and separate the self-propelled particles.

In this article, 
 we numerically study 
the transport properties and distribution of chiral active Brownian particles in a channel with the Poiseuille flow. 
For example, the average particle velocity and effective diffusion coefficient, of both nonchiral and chiral self-propelled particles in a two-dimensional microchannel flow.
The flow profile is prescribed by a Poiseuille flow, and the collisional dynamics of the particles at the channel walls are modeled by sliding reflecting boundary conditions. 
We focus on finding how the self-propulsion mechanism and strength of the fluid flow influence the transport characteristics in this microchannel flow. 
The rest of this article is organized as follows.
In section -\ref{Model}, we introduce the model for the self-propelled particles in a two-dimensional microchannel with a Poiseuille flow. 
The transport characteristics for both the nonchiral and chiral particles are investigated in section -\ref{Nonchiral} and section -\ref{Chiral}, respectively.
Finally, we present the main conclusions in section -\ref{Conclusions}.

\section{Model}\label{Model}

\begin{figure}[htb!]
\centering
\includegraphics[scale = 0.75]{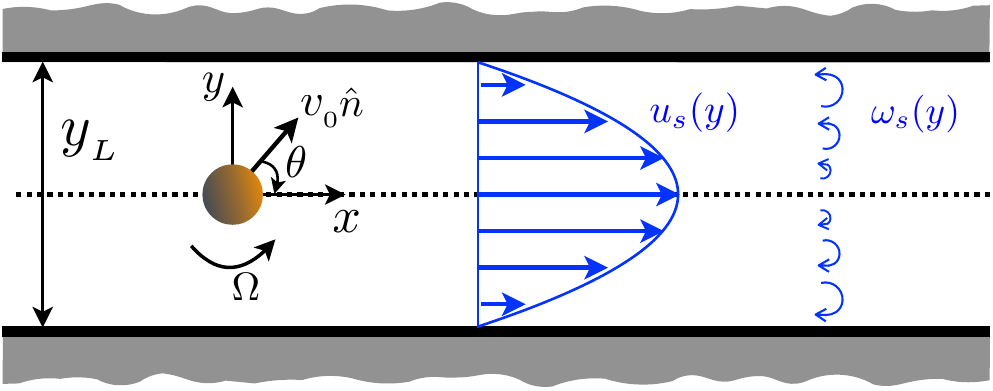}
\caption{Schematic illustration of the active Brownian particle in a two-dimensional microchannel with a Poiseuille flow $u_s(y)$. The self-propelled velocity $v_0$, self-propelled angle $\theta$, angular velocity $\Omega$, local shear rate $\omega_s (y)$, and width of the channel $y_L$ are indicated. 
The sliding reflecting boundary conditions at the channel walls assure the confinement of self-propelled particle inside the channel.}
\label{fig:chw}
\end{figure}

We consider active Brownian particles suspended in a two-dimensional microchannel where a Poiseuille flow is imposed (see Fig.~\ref{fig:chw}). 
The particle possesses an orientational degree of freedom characterized by an unit vector $\hat{n} = (\cos \theta, \sin \theta)$, where $\theta$ is the angle relative to the $x-$axis.
This particle self-propels along its orientation with the self-propelled velocity $\vec{v}_0 = v_0 \hat{n}$ with constant modulus $v_0$. 
If this particle is chiral, $\theta$ is subjected to a rotation with angular velocity $\Omega$ as a consequence of torque acting on the particle \cite{Lowen_RMP,Volpe_AJP}.
Additionally, the two-dimensional Poiseuille flow $u_s (y)$ directed along the $x-$axis with a shear gradient along the $y-$axis affects the dynamics of this particle by forcing its self-propulsion velocity to rotate under the action of the local shear rate $\omega_s (y) = -(1/2) du_s (y)/ dy$.
In the overdamped limit, the equation of motion of the particle reads \cite{Hagen_PRE}: 
\begin{subequations}
\begin{align}
\frac{dx}{dt} &= u_s (y) + v_0 \cos \theta + \sqrt{2 D_0}~ \xi_x (t), \\
\frac{dy}{dt} &= v_0 \sin \theta + \sqrt{2 D_0}~ \xi_y (t), \\
\frac{d \theta}{dt} &= \omega_s (y) + \Omega +  \sqrt{2 D_\theta} ~\xi_\theta (t),
\end{align}
\label{eq:Langevin} 
\end{subequations}
where $\vec{r} = (x, y)$ is the position of the particle, and $D_0$ and $D_\theta$ are the translational and rotational diffusion constants, respectively.
 $\xi_x (t)$, $\xi_y (t)$, and $\xi_\theta (t)$ are the Gaussian white noises satisfying $\langle \xi_i (t) \rangle = 0$ and $\langle \xi_i (t) \xi_j (t') \rangle = \delta_{i,j} \delta(t - t')$ for $i, j = x, y, \theta$.
 Note that the direction of the particle randomly varies with the time scale $\tau_\theta = 2/D_\theta$. 
 Accordingly, the trajectory of the particle approximately combines the self-propulsion length $l_\theta = v_0 \tau_\theta$ and a circular arc of radius $R_\Omega = v_0/|\Omega|$ \cite{Lowen_RMP, Teeffelen_PRE}.
 
 The two-dimensional microchannel is described by two parallel walls with a separation distance $y_L$, see Fig.~\ref{fig:chw}. 
 For the Poiseuille flow between the channel walls, the flow velocity is prescribed as 
 \begin{equation}
 u_s(y) = u_0 \left[1 -  \left( \frac{2y}{y_L} \right)^2 \right],
 \label{eq:flow_field}
 \end{equation}
 where $u_0$ is the maximum flow speed at the center of the channel and the corresponding vorticity is $\omega_s (y) = 4 u_0 y/ y_L ^2$.
 Although the flow velocity decreases away from the centerline and vanishes at the channel walls $y = \pm y_L/2$, the local shear rate increases linearly with $y$, see Fig.~\ref{fig:chw}.  
 In order to have a dimensionless description, we henceforth scale all length variables by the width of the channel $y_{L}$, i.e., $x \to x/y_{L}$ and $y \to y/y_{L}$.
Analogously, we rescale the time $t \to v_0 t/y_{L}$, so as to work with a constant rescaled self-propelled velocity $v_0 = 1$.
 The other rescaled parameters read $D_0 \to D_0/(y_{L} v_0)$, $D_{\theta} \to y_{L} D_{\theta}/v_0$, $u_0 \to u_0/v_0$, and $\Omega \to y_{L} \Omega/v_0$.
 In the rest of the article, we use dimensionless variables only.
  The collisional dynamics of the particle at the channel walls is modeled as follows. 
 The translational velocity $\dot{\vec{r}}$ is elastically reflected \cite{Khatri_JCP, Khatri_pre1, Khatri_JSTAT, Khatri_pre2}, and the coordinate $\theta$ is unchanged during the collision (sliding reflecting boundary condition \cite{Ghosh_PRL, Zhang_et_al@PF:2010}).
 Accordingly, the particle slides along the walls for an average time of the order of $\tau_\theta$, until the fluctuations in $\theta$, i.e., $\xi_\theta (t)$, redirect it towards the interior of the channel. 
 Note that on modeling the boundary conditions, we assume that particles are small in size such that we can safely neglect 
 the hydrodynamic interaction between the particles and the channel walls \cite{Lauga_JFM}.

The key quantities of interest are the average velocity and effective diffusion coefficient of the particles. 
Since the Langevin equations (\ref{eq:Langevin}) turn nonlinear in the presence of the Poiseuille flow, the explicit analytical expressions of the average velocity and effective diffusion coefficient cannot be obtained.
For this reason, the behavior of these quantities can be obtained using Brownian dynamics simulations performed by the integration of Langevin equations (\ref{eq:Langevin}) using standard stochastic Euler algorithm over $2 \times 10^4$ trajectories with sliding reflecting boundary conditions at the channel walls.
 As initial conditions, we have assumed that at $t = 0$, all the particles are randomly distributed with random orientations in the channel.
 Since particles along the $y-$direction are confined, we only calculate the average velocity and effective diffusion coefficient along the $x-$direction. 
 Numerically, the average velocity and effective diffusion coefficient along the $x-$direction are, respectively, calculated as 
 \begin{align}
 v &= \lim_{t\to\infty} \frac{\langle x(t) \rangle}{t}, \label{eq:velocity}\\
 D_{eff}  &= \lim_{t\to\infty} \frac{\langle x^2(t) \rangle - \langle x(t) \rangle ^2 }{2 \,t}. \label{eq:diffusion}
 \end{align}
 
 \section{Diffusion of nonchiral particles}\label{Nonchiral}
 
\begin{figure}[htb!]
\centering
\includegraphics[scale = 1.0]{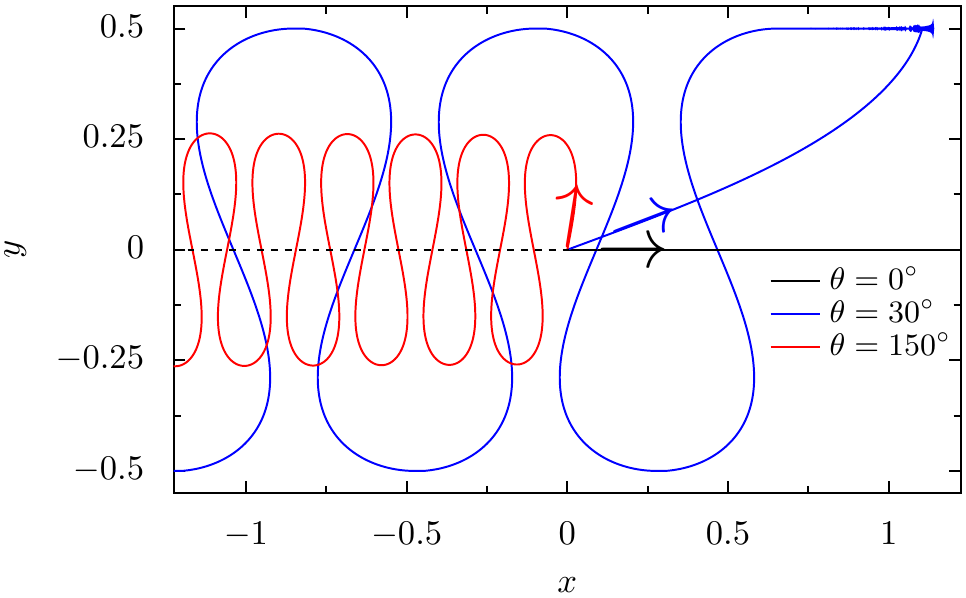}
\caption{Deterministic trajectories of a nonchiral ($\Omega = 0$) particle in 
a two-dimensional microchannel with a Poiseuille flow prescribed by equation~(\ref{eq:flow_field}). The particle starts at the center of the channel, i.e., $x = y = 0$, with different values of the self-propelled angle $\theta$. Trajectories are obtained by integrating equations~(\ref{eq:Langevin}) numerically for $D_0 = D_\theta = 0$ and $u_0 = 1$.}
\label{fig:graph2}
\end{figure} 

\begin{figure}[t]
\centering
\includegraphics[scale = 0.85]{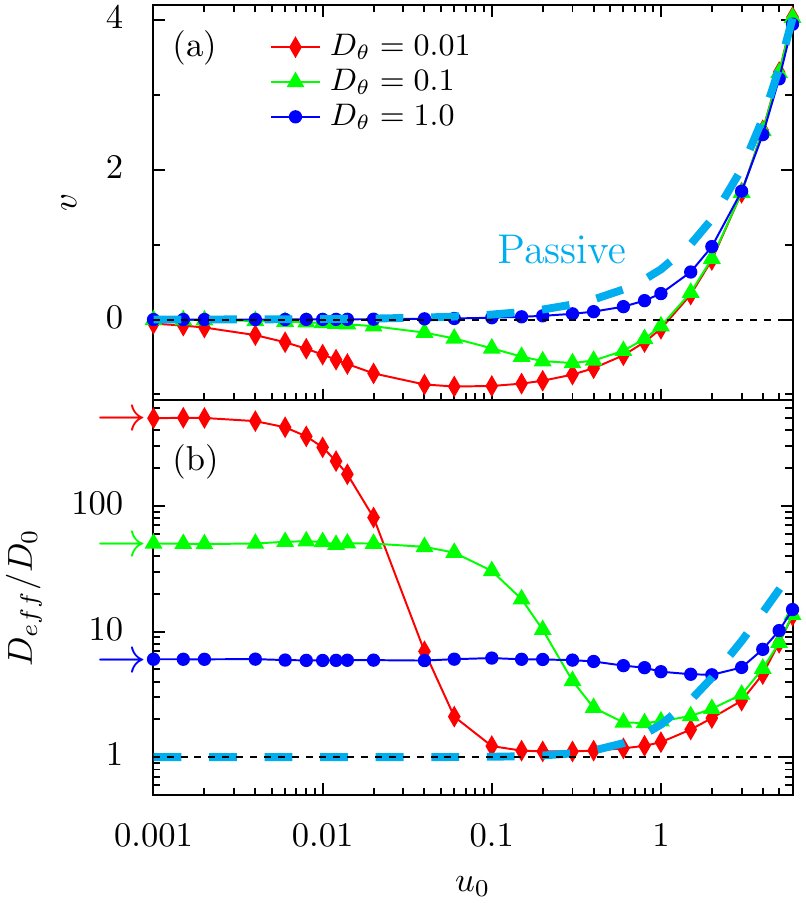}
\caption{The average velocity $v$ as a function of the flow strength $u_0$ is depicted in (a) for different values of rotational diffusion rate $D_\theta$. 
The corresponding scaled effective diffusion coefficient $D_{eff}/D_0$ is depicted in (b). 
The right arrows denote the values of $D_{eff}/D_0$ in the absence of Poiseuille flow calculated from the well known analytical expression  $D_{eff} = D_0 + \tau_\theta/4$ \cite{Howse_PRL}. The solid line is a guide to the eyes. The set parameters are $D_0 = 0.1$ and $\Omega = 0$. The dashed cyan line corresponds to the passive Brownian particles.}
\label{fig:graph3}
\end{figure}
 
Although the transport properties of nonchiral active Brownian particles in a channel with the Poiseuille flow are recently reported  by  Peng and Brady \cite{Peng_PRF}; however, for completeness, in this section, we present 
the diffusive transport of nonchiral particles in a Poiseuille flow.
To analyze the main effects of Poiseuille flow on the nonchiral ($\Omega = 0$) particle  dynamics, in Figure~\ref{fig:graph2}, we show the deterministic ($D_0 = D_\theta = 0$) trajectories of a particle positioned initially at the center of the channel with different values of the self-propelled angle $\theta$. 
When $\theta = 0^{\circ}$, the particle continues to drift along the centerline, i.e., along the flow direction without any rotation ($\omega_s(y) = 0$). 
However, when $0^{\circ} < \theta \leq \theta_c$ ($\theta_c \approx 120^{\circ}$), the particle can reach the upper wall and then drifts in the opposite direction to the fluid flow (upstream drift). 
It is because, near the upper wall, the fluid velocity tends to zero, and due to the large fluid vorticity, the particle reoriented counterclockwise (see Fig.~\ref{fig:chw}). 
The particle that exhibits upstream drift will then reach the lower wall and reorient clockwise.
This leads to a periodic motion.
Note that within this range of $\theta$, the qualitative behavior of all trajectories remains the same. 
In particular, the value of $\theta_c$ depends on the flow strength $u_0$.
Interestingly, when $\theta$ is greater than the critical self-propelled angle $\theta_c$, the particle would not reach the walls (see Fig.~\ref{fig:graph2}). 
However, it still performs a periodic motion, where the temporal period is much longer and the spatial period is shorter. 
Therefore, we obtain a confined trajectory, where walls do not cause the confinement but due to the combined effect of Poiseuille flow and self-propulsion mechanism.
For $\theta = 180^{\circ}$, the particle would not move at the middle of the channel because, in this situation, $v_0 = u_0$.
Similar trajectories are reported in references \cite{Peng_PRF, Zottl_Stark@PRL:2012, Costanzo_JPCM}.

\begin{figure}[t]
\centering
\includegraphics[scale = 0.85]{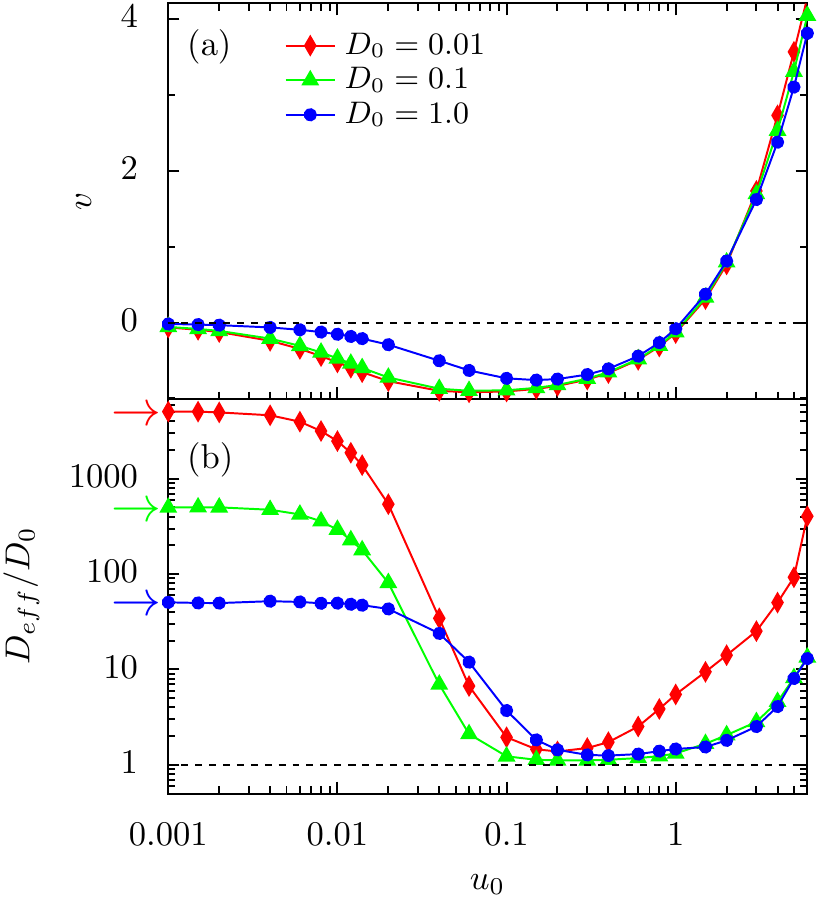}
\caption{The average velocity $v$ as a function of the flow strength $u_0$ is depicted in (a) for different values of translational diffusion constant $D_0$. 
The corresponding scaled effective diffusion coefficient $D_{eff}/D_0$ is depicted in (b). 
The right arrows denote the values of $D_{eff}/D_0$ in the absence of Poiseuille flow calculated from the well known analytical expression $D_{eff} = D_0 +  \tau_\theta/4$.
The solid line is a guide to the eyes. The set parameters are $D_\theta = 0.01$ and $\Omega = 0$.}
\label{fig:graph4}
\end{figure} 

Figure~\ref{fig:graph3} depicts the average velocity $v$ and effective diffusion coefficient $D_{eff}/D_0$ as a function of the flow strength $u_0$ for different values of rotational diffusion rate $D_\theta$.
In the limit $u_0 \ll v_0$, the fluid velocity can be neglected, and $v$ tends to zero and $D_{eff}$ agrees with the well known analytical expression $D_{eff} = D_0 + \tau_\theta/4$ \cite{Howse_PRL} for different values of $D_\theta$.
Interestingly, the particles exhibit upstream drift for the small and moderate values of $D_\theta$, when $u_0 \le v_0$, as demonstrated in Figure~\ref{fig:graph2}.
Therefore, $v$ becomes negative.
Note that this upstream drift is only possible for active particles, whereas passive particles would simply follow the fluid velocity profile. 
For the higher $D_\theta$ value, i.e., $D_\theta \rightarrow \infty$, the self-propelled angle $\theta$ changes rapidly; therefore, the motion of active particles is similar to passive Brownian motion with a positive $v$ for any strength of $u_0$ (see Fig.~\ref{fig:graph3}(a)). 
As one would expect, the magnitude of $v$ decreases with increasing $D_\theta$. 
It is worth to point out that the combined effect of the Poiseuille flow and self-propulsion mechanism suppresses $D_{eff}$. 
When $u_0 \gg v_0$, the fluid velocity at the middle of the channel dominates over the self-propulsion mechanism; thus, in this regime, both $v$ and $D_{eff}$ approach to the transport properties of passive particles. 
Consequently, $v$ is positive for any value of $D_\theta$, and both $v$ and $D_{eff}$ increase with increasing $u_0$.
The similar nonmonotonic behavior of both $v$ and $D_{eff}$ as a function of $u_0$ is recently reported by Peng and Brady \cite{Peng_PRF} in the context of nonchiral active Brownian particles in a channel with the Poiseuille flow. 

Figure~\ref{fig:graph4} shows the average velocity $v$ and effective diffusion coefficient $D_{eff}/D_0$ as a function of the flow strength $u_0$ for different values of translational diffusion constant $D_0$.
We can see that the qualitative behavior of $v$ and $D_{eff}$, which has been explained earlier, does not change for different values of $D_0$. 
Also, $v$ cannot be reversed for any value of $D_0$. 
However, as one would expect, $v$ decreases with increasing $D_0$ (see Fig.~\ref{fig:graph4}(a)).
Particularly, in the limit $u_0 \ll v_0$, as mentioned earlier, $v$ tends to zero and $D_{eff}$ agrees with the well known analytical expression $D_{eff} = D_0 + \tau_\theta/4$ for different values of $D_0$.
In the other limit, i.e., $u_0 \gg v_0$, the fluid velocity dominates over the self-propulsion mechanism; thus, in this regime, $v$ is positive, and both $v$ and $D_{eff}$ increase with increasing $u_0$.

\section{Diffusion of chiral particles}\label{Chiral} 

\begin{figure}[htb!]
\centering
\includegraphics[scale = 1.0]{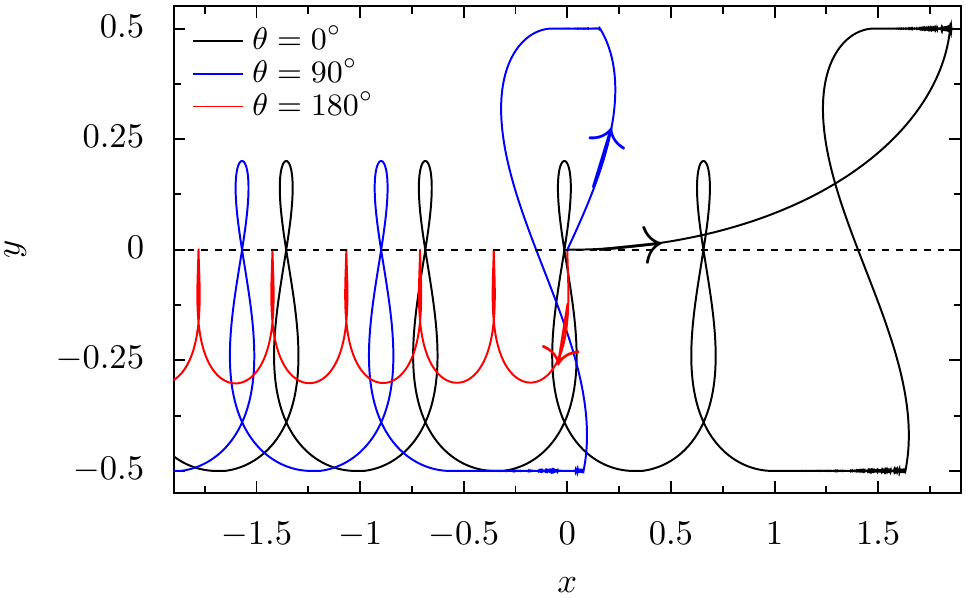}
\caption{Deterministic trajectories of a chiral particle in the two-dimensional microchannel with the Poiseuille flow prescribed by equation~(\ref{eq:flow_field}). The particle starts at the center of the channel, i.e., $x = y = 0$, with different values of the self-propelled angle $\theta$. Trajectories have been produced by integrating equations~(\ref{eq:Langevin}) for $D_0 = D_\theta = 0$, $\Omega = 0.3$, and $u_0 = 1$.}
\label{fig:graph5}
\end{figure}

In this section, we study the diffusive transport of chiral particles ($\Omega \neq 0$). 
First, in order to analyze the main effects of Poiseuille flow on the chiral particle dynamics, in Figure~\ref{fig:graph5}, we study the deterministic trajectories ($D_0 = D_{\theta} = 0$) of a particle positioned initially at the center of the channel with different values of the self-propelled angle $\theta$ and for $\Omega > 0$.
When $0^{\circ} \leq \theta \leq \theta_c$ ($\theta_c \approx 145^{\circ}$), the particle exhibits an upstream drift, and its trajectory initially has a transient and then a steady state part.
In the latter, the particle performs a periodic motion.
As one would expect, due to the chirality of the particle, the up and down symmetry in the motion is broken; thus, in the steady state part, the particle would not reach the upper channel wall.
If one considers the case $\Omega < 0$,  an opposite behavior can be observed. 
Note that within this range of $\theta$, the qualitative behavior of all the trajectories remains the same. 
In particular, the value of $\theta_c$ depends on the flow strength $u_0$ and angular velocity $\Omega$. 
Interestingly, when $\theta > \theta_c$, the initial transient part in the trajectory disappears, and the particle would not reach the walls (see Fig.~\ref{fig:graph5}). 
The particle still performs a periodic motion, where the temporal period is much longer and the spatial period is shorter. 
Consequently, we obtain a confined trajectory caused by the Poiseuille flow and self-propulsion mechanism rather than the channel walls.

\begin{figure}[t]
\centering
\includegraphics[scale = 0.85]{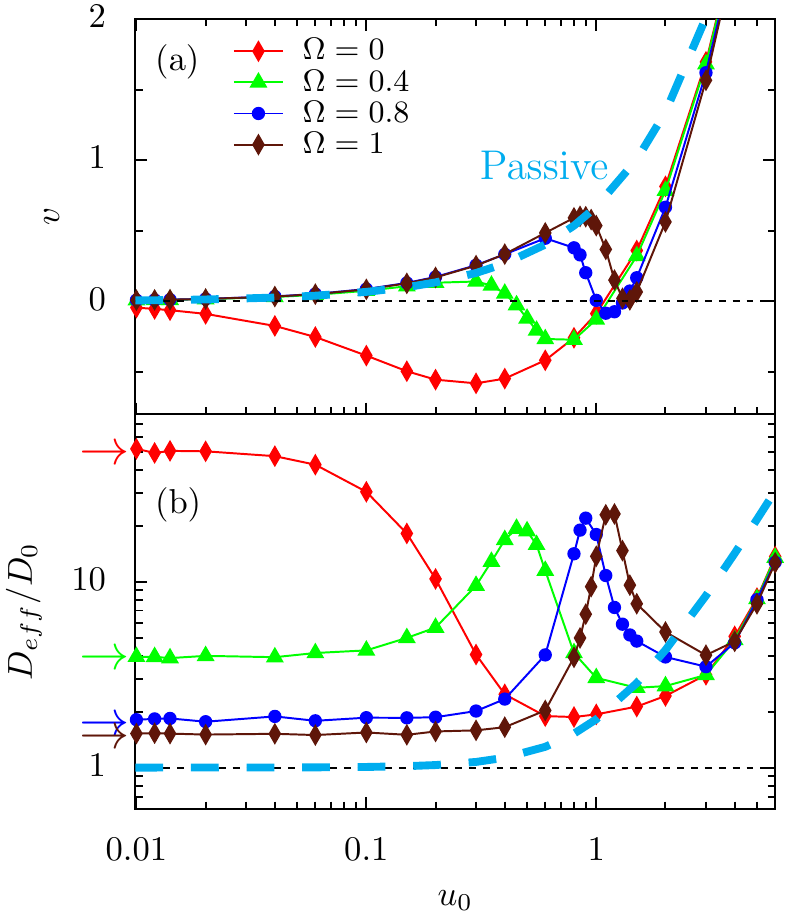}
\caption{The average velocity $v$ as a function of the flow strength $u_0$ is depicted in (a) for different values of angular velocity $\Omega$. 
The corresponding scaled effective diffusion coefficient $D_{eff}/D_0$ is depicted in (b). 
The right arrows denote the values of $D_{eff}/D_0$ in the absence of Poiseuille flow calculated from the well known analytical expression $D_{eff} = D_0 \, + \, (\tau_\theta/4)/[1 \, + \, (\Omega \tau_\theta/2)^2]$ \cite{Teeffelen_PRE, Ebbens_PRE}.  The solid line is a guide to the eyes.
The set parameters are $D_\theta = D_0 = 0.1$.
The dashed cyan line corresponds to the passive Brownian particles.}
\label{fig:graph6}
\end{figure}

Figure~\ref{fig:graph6} shows the average velocity $v$ and the scaled effective diffusion coefficient $D_{eff}/D_0$ as a function of the flow strength $u_0$ for different values of angular velocity $\Omega$. 
As before, in the limit $u_0 \ll v_0$, the fluid velocity can be neglected; thus, $v \to 0$ and $D_{eff}$ agrees with the well known analytical expression  $D_{eff} = D_0 + (\tau_\theta/4)/[1 + (\Omega \tau_\theta/2)^2]$ \cite{Teeffelen_PRE, Ebbens_PRE} for different values of $\Omega$.
For $\Omega \le 0.8$, it is found that $v$ reversal occurs as a function of $u_0$.
It indicates that there exist finite values of $u_0$ for which the particles exhibit upstream drift. 
On the other hand, for $\Omega > 0.8$, the self-propelled angle $\theta$ changes rapidly; thus, $v$ remains positive for any value of $u_0$. 
Interestingly, when $\Omega$ is comparable to the local shear rate due to the Poiseuille flow $\omega_s(y)$  near the lower channel wall, $\Omega$ cancels out the effect of $\omega_s(y)$; thus, the particles aggregate near the lower wall. 
As a result, $v$ suddenly drops and tends to zero, and $D_{eff}$ exhibits an enhanced peak. 
Note that if one considers that $\Omega < 0$, the same would happen on the upper channel wall. 
In the noiseless limit, this happens for $\Omega \simeq \omega_s(-y_L/2)$. 
This condition can be regarded as the onset of the aggregation mechanism.
As mentioned earlier, in the regime $u_0 \gg v_0$, both $v$ and $D_{eff}$ approach to the transport properties of passive particles. 
Consequently, $v$ and $D_{eff}$ increase with increasing $u_0$.

\begin{figure}[t]
\centering
\includegraphics[scale = 1.0]{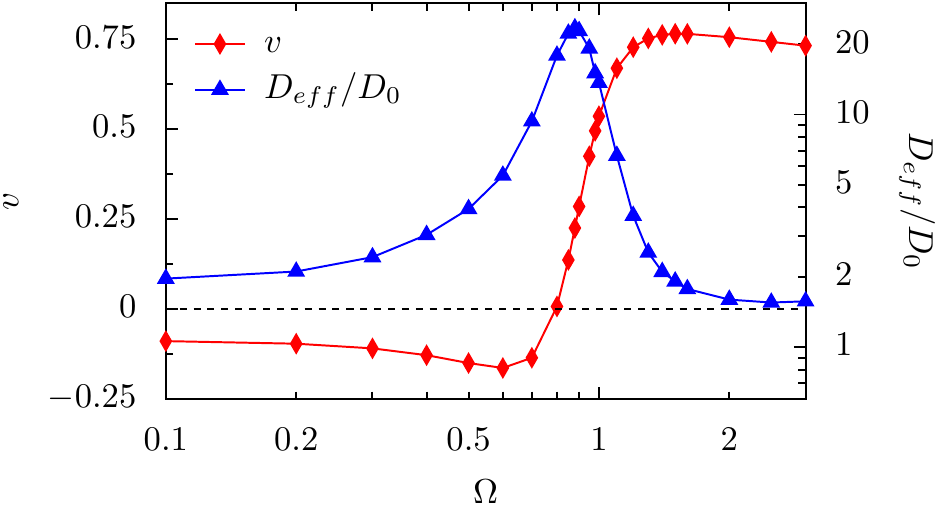}
\caption{The average velocity $v$ and the scaled effective diffusion coefficient $D_{eff}/D_0$ as a function of the angular velocity $\Omega$ for a fixed value of the flow strength $u_0 = 1$.  The solid line is a guide to the eyes. The set parameters are $D_\theta = D_0 = 0.1$.}
\label{fig:graph7}
\end{figure}

Figure~\ref{fig:graph7} depicts the average velocity $v$ and the scaled  effective diffusion coefficient $D_{eff}/D_0$ as a function of the angular velocity $\Omega$ for a fixed value of the flow strength $u_0 = 1$.
For $\Omega < 0.8$, the particles exhibit upstream drift resulting in the negative $v$. 
On further increasing $\Omega$, the self-propelled angle $\theta$ changes rapidly; thus, the average velocity reverses its direction and becomes positive. 
Interestingly, as discussed earlier, when $\Omega$ is comparable to the local shear rate $\omega_s(y)$ near the lower channel wall, $\Omega$ cancels out the effect of $\omega_s(y)$; therefore, the particles aggregate near the lower wall. 
In the absence of noise, for the considered $u_0 = 1$, this would happen for $\Omega \simeq 1$. 
Whereas, in the presence of noise, this happens for $\Omega \simeq 0.8$. 
Accordingly, $v \to 0$ and $D_{eff}$ exhibits an enhanced peak.

\begin{figure*}[hbt!]
\centering
\includegraphics[scale = 0.9]{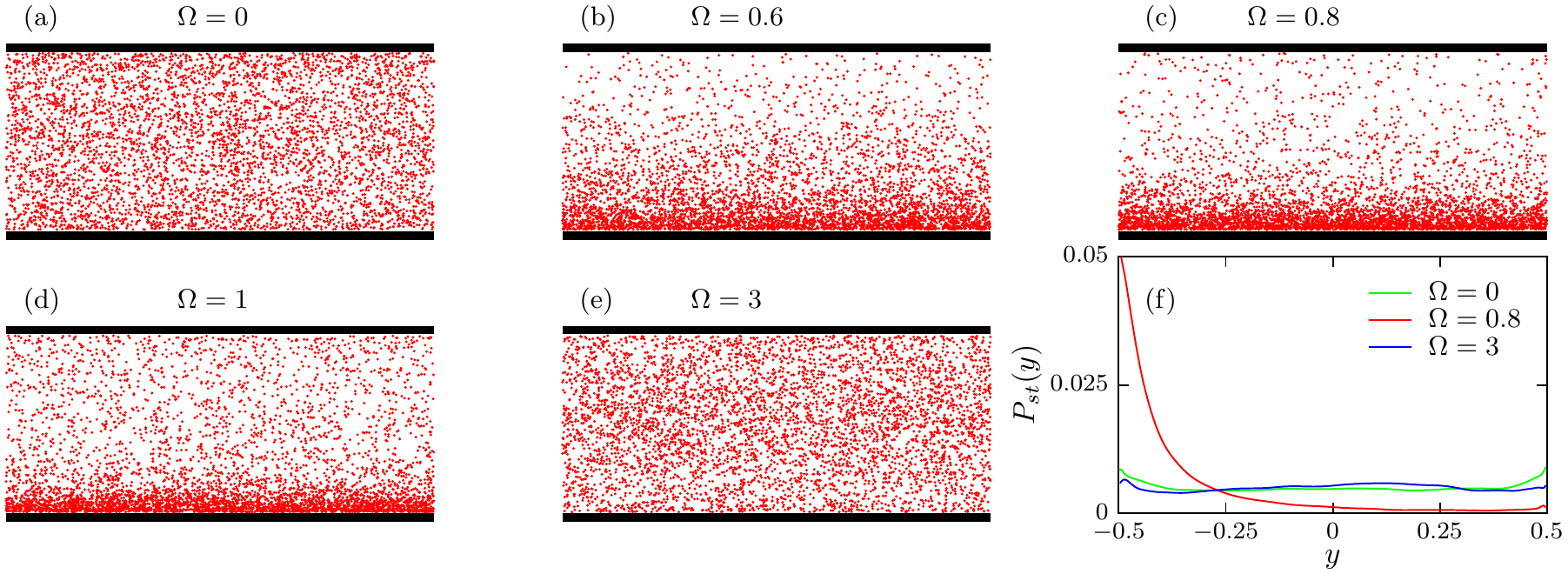}
\caption{The steady state distribution of chiral particles, for various values of the angular velocity $\Omega$, is depicted in (a)-(e). 
The corresponding normalized probability distributions $P_{st}(y)$ along the $y-$direction are depicted in (f).	
The set parameters are $D_\theta = D_0 = 0.1$ and $u_0 = 1$.}
\label{fig:graph8}
\end{figure*}

Figure~\ref{fig:graph8} depicts the steady state distribution and the corresponding normalized probability distribution $P_{st}(y)$ of chiral particles for various values of the angular velocity $\Omega$. 
For $\Omega \rightarrow 0$, the particles distribute uniformly both in the upper and lower regions of the channel. 
This is reflected in $P_{st}(y)$ (see Fig.~\ref{fig:graph8}(f)). 
Interestingly, on further increasing $\Omega$, the particles start to aggregate near the lower channel wall, and the aggregation is maximum for $\Omega \simeq 0.8$.
This onset of aggregation mechanism is due to the fact that when $\Omega$ is comparable to the local shear rate $\omega_s(y)$ near the lower channel wall, $\Omega$ cancels out the effect of $\omega_s(y)$.
Accordingly, $P_{st}(y)$ is maximum near the lower wall, and it monotonically decreases away from the lower wall.
As mentioned earlier, if one considers the negative sign of $\Omega$, the same would happen on the upper channel wall. 
When $\Omega$ dominates over $\omega_s(y)$, as one would expect, the particles distribution and the corresponding probability distribution in both the upper region and bottom region of the channel become the same (see (e) and (f) in Fig.~\ref{fig:graph8}).

\section{Conclusions}\label{Conclusions}

In this work, we have numerically studied the diffusive transport of both nonchiral and chiral self-propelled particles in a two-dimensional microchannel with a Poiseuille flow. 
Using Brownian dynamics simulations, we have investigated that the transport characteristics, the average velocity and effective diffusion coefficient, strongly depend on the self-propulsion mechanism and the flow strength. 
We have demonstrated that the particles exhibit upstream drift, which is a signature of the active particles. 
It is found that the direction of the average velocity can be reversed by suitably tuning the rotational diffusion rate, flow strength, and angular velocity. 
In addition, for the case of nonchiral particles, the effective diffusion coefficient is suppressed by the presence of the Poiseuille flow. 
Interestingly, for the case of chiral particles, we have shown that the particles aggregate near the lower channel wall, and correspondingly the effective diffusion coefficient exhibits an enhanced peak. 
The present study can provide insights into flow controlled diffusive behavior of self-propelled particles in the microchannel flow.
Also, the results are expected to be instructive to design lab-on-a-chip devices for separating the living and non-living organisms as well as chiral and nonchiral self-propelled particles \cite{Chin_Lab, Yang_SM}.

\section{Acknowledge}
This work was supported by the Indian Institute of Technology Kharagpur.

\end{document}